\documentclass[12pt]{article}
\pdfoutput=1
\usepackage{geometry,enumerate,amsmath,amssymb,amsfonts}
\usepackage{fullpage}
\usepackage{graphicx}
\usepackage{caption}
\usepackage{subcaption}
\numberwithin{equation}{section}
\newcommand{\be}{\begin{equation}}
\newcommand{\ee}{\end{equation}}
\newcommand{\bea}{\begin{eqnarray}}
\newcommand{\eea}{\end{eqnarray}}
\newcommand{\bphi}{\mbox{\boldmath $\phi$}}

\renewcommand{\epsilon}{\varepsilon}

\begin{document}
\title{
\begin{flushright}\ \vskip -2cm {\small{\em DCPT-14/41}}\end{flushright}
\vskip 2cm 
Aloof Baby Skyrmions}
\author{\ \\ Petja Salmi and Paul Sutcliffe\\[10pt]
{\em \normalsize Department of Mathematical Sciences,
Durham University, Durham DH1 3LE, U.K.}\\[10pt]
{\normalsize 
petja.salmi@durham.ac.uk \quad\&\quad\  p.m.sutcliffe@durham.ac.uk}
}
\date{September 2014}
\maketitle
\begin{abstract}

We show that a suitable choice for the potential term in the
two-dimensional baby Skyrme model yields solitons that have a 
short-range repulsion and a long-range attraction. The solitons are therefore
aloof, in the sense that static multi-soliton bound states
have constituents that preserve their individual identities and are 
sufficiently far apart that tail interactions yield small binding 
energies. The static multi-soliton solutions are found to have a cluster 
structure that is reproduced by a simple binary species particle model. 
In the standard three-dimensional Skyrme model of nuclei,
solitons are too tightly bound and are often too symmetric, due to
symmetry enhancement as solitons coalesce to form bound states.
The aloof baby Skyrmion results endorse a way to resolve these issues
and provides motivation for a detailed study of the related 
three-dimensional version of the Skyrme model. 
\end{abstract}

\newpage 
\section{Introduction}\quad
In the standard three-dimensional Skyrme model of nuclei, solitons
coalesce to form bound states that are often highly symmetric \cite{BS3}.
However, in comparison to experimental data on nuclei, the solitons are too tightly bound
and are often too symmetric. Including massive pions does change the 
soliton symmetries (for a sufficiently large number of solitons) and 
produces solutions with a more realistic cluster structure \cite{BMS}, but
the solitons are still too tightly bound. One strategy for reducing 
soliton binding energies is to consider a modified Skyrme model that is a 
 perturbation from a BPS theory, either by including
terms in the Lagrangian that are only zero or sixth order in derivatives
\cite{ASW,ANSW} or by adding a tower of vector mesons 
to the standard Skyrme model \cite{Su}.  

An alternative approach to a near BPS theory is to consider a
family of theories that interpolates between
the standard Skyrme model and a theory in which inter-soliton forces are 
repulsive. Binding energies can then be made suitably small by choosing the 
value of the interpolation parameter sufficiently close to the repulsive
limit. An appropriate repulsive theory in three 
dimensions has recently been identified
and consists of a Skyrme term and a potential term,
 chosen so that a topological energy bound can be attained only in the
 one-soliton sector \cite{Ha}. This makes the above interpolation 
scheme feasible and work is currently underway to investigate this in detail
\cite{GHS}.   
 
The purpose of the current paper is to investigate a lower-dimensional 
analogue of this scheme, where both numerical and analytic methods 
are easier to employ. The standard baby Skyrme model \cite{PSZ} 
is a two-dimensional analogue of the standard Skyrme model. 
Furthermore, a version of the baby Skyrme model with repulsive solitons 
has been known for a long time \cite{LPZ}, and in fact this has the same
potential term as in the recently discovered repulsive Skyrme model \cite{Ha}.
We demonstrate that the interpolating family includes theories
where the solitons have a short-range repulsion and a long-range 
attraction. These solitons are therefore
aloof, in the sense that static multi-soliton bound states
have constituents that preserve their individual identities and are 
sufficiently far apart that tail interactions yield small binding 
energies. The static multi-soliton solutions are found to have a cluster 
structure that is reproduced by a simple binary species particle model. 

The results presented on two-dimensional aloof solitons 
provides evidence to support the view that similar aloof solitons
in three dimensions could address some of the current problematic issues
with soliton binding energies and symmetries in the standard Skyrme model. 
Not only does this work supply additional motivation to study the 
three-dimensional problem but it also demonstrates the utility of point
particle models in predicting the forms and energies
 of multi-soliton bound states. 

\section{The aloof baby Skyrme model}\quad
The field of the baby Skyrme model is a three-component unit
vector ${\bphi}=(\phi_1,\phi_2,\phi_3)$, with associated static
energy 
\be
E=\int \bigg(\frac{1}{2}\partial_i\bphi\cdot\partial_i\bphi
+\frac{1}{4}(\partial_i\bphi\times\partial_j\bphi)\cdot
(\partial_i\bphi\times\partial_j\bphi)+V\bigg)\,d^2x,
\label{energy}
\ee
where $V(\bphi)$ is a potential and $x_i$, 
are coordinates in the plane, with $i=1,2.$  
In the standard baby Skyrme model \cite{PSZ} 
the potential is taken to be 
\be
V=m^2(1-\phi_3),
\label{pot1}
\ee
which is the analogue of the conventional pion mass term
in the Skyrme model \cite{AN}. The constant $m$ gives the
mass of the fields $\phi_1$ and $\phi_2,$ associated with
elementary excitations around the unique vacuum $\bphi=(0,0,1).$

Finite energy requires that the field takes the vacuum value
at all points at spatial infinity, $\bphi(\infty)=(0,0,1).$
This compactification means that
topologically $\bphi$ is a map between two-spheres, with an
associated integer winding number $N\in\mathbb{Z}=\pi_2(S^2).$
This topological charge (or soliton number) is the analogue of the
baryon number in the Skyrme model and may be calculated as
\be
N=-\frac{1}{4\pi}\int \bphi\cdot(\partial_1\bphi\times\partial_2\bphi)\, d^2x.
\label{charge}
\ee
There is an attractive channel for inter-soliton forces in the 
standard baby Skyrme model and hence there
are multi-soliton bound states that have been investigated numerically in
some detail \cite{PSZ,Fo}. In particular, the 2-soliton bound state is
radially symmetric as there is no short-range repulsion in the attractive
channel, so two single solitons coalesce into one radial structure.

Several other choices for the potential have been investigated, but the one
of particular relevance to the present paper is the potential introduced in 
\cite{LPZ}
\be
V=m^2(1-\phi_3)^4.
\label{pot2}
\ee
The fields $\phi_1$ and $\phi_2$ are both massless for this choice of potential,
so $m$ is simply a parameter of the theory 
that controls the size of the soliton, rather than a mass. 
The theory with potential (\ref{pot2}) has the interesting property that the
$N=1$ soliton can be written down explicitly in closed form, although there
are no multi-soliton bound states as there is a repulsive force between
solitons \cite{LPZ}. 
These features are explained by the fact that there is a topological
energy bound \cite{IRPZ}
\be
E\ge 4\pi N\bigg(1+\frac{4\sqrt{2}}{3}m\bigg)
\ee
that is attained only in the $N=1$ sector, by the 1-soliton solution
\be
\bphi=\bigg(\frac{2\lambda x_1}{x_1^2+x_2^2+\lambda^2},
\frac{2\lambda x_2}{x_1^2+x_2^2+\lambda^2},
\frac{x_1^2+x_2^2-\lambda^2}{x_1^2+x_2^2+\lambda^2}
\bigg),
\ee 
where $\lambda=1/(2^{1/4}\sqrt{m})$ is the width of the soliton.
With the repulsive potential (\ref{pot2}) 
this is the exact 1-soliton solution
positioned at the origin $(x_1,x_2)=(0,0)$, where $\bphi=(0,0,-1)$,
but the position can be moved to an arbitrary point in the plane by making use
of the translational symmetry of the theory. An arbitrary phase can also be
introduced into the solution by acting with the global $U(1)$ symmetry that
rotates the first two components of $\bphi.$  

Given the potentials (\ref{pot1}) and (\ref{pot2}), that produce an attractive 
and repulsive force between solitons respectively, we can
now implement the strategy described in the introduction and consider a 
potential that is a linear combination 
of the two, with the aim of generating a short-range repulsion and a 
long-range attraction that combine to yield weakly bound multi-solitons.
Explicitly, the potential is taken to be
\be
V=m^2\bigg(\mu(1-\phi_3)+(1-\mu)(1-\phi_3)^4\bigg),
\label{pot3}
\ee
where $\mu\in[0,1]$ is a parameter that interpolates
between the repulsive model $\mu=0$ and the standard baby Skyrme model $\mu=1.$
It will turn out that the choice $\mu=\frac{1}{2}$ is sufficient to achieve our intended aims, so we restrict to this value for most of our investigations 
and concentrate on the resulting potential
\be
V=\frac{m^2}{2}(1-\phi_3)(1+(1-\phi_3)^3).
\label{pot4}
\ee
The value of the constant $m$ merely determines an overall length scale,
with the mass of the elementary excitations equal to ${m}/{\sqrt{2}}.$
The numerical studies in \cite{PSZ,Fo}, for the potential
(\ref{pot1}), selected the value $m^2=0.1,$ so to aid comparison
with previous studies we take this same value for $m.$ 

\section{Soliton solutions}\quad
In this section we investigate static soliton solutions of the aloof
baby Skyrme model with potential (\ref{pot4}) and $m^2=0.1.$ 
A radially symmetric field with topological charge $N$ is given by
\be
\bphi=(\sin f\cos(N\theta+\chi),\sin f\sin(N\theta+\chi), \cos f),
\label{radial}
\ee
where $r$ and $\theta$ are polar coordinates in the plane
with $f(r)$ a monotonically decreasing radial profile function
such that $f(0)=\pi$ and $f(\infty)=0.$ 
The constant $\chi$ is an internal phase that can be set to zero 
by applying the global symmetry that rotates the $\phi_1,\phi_2$ components.
Although the internal phase is irrelevant in the above radial ansatz, the
relative phase between two separated solitons is important and plays a 
crucial role in the inter-soliton force.  

\begin{figure}[ht]\begin{center}
\hbox{
\includegraphics[width=7cm]{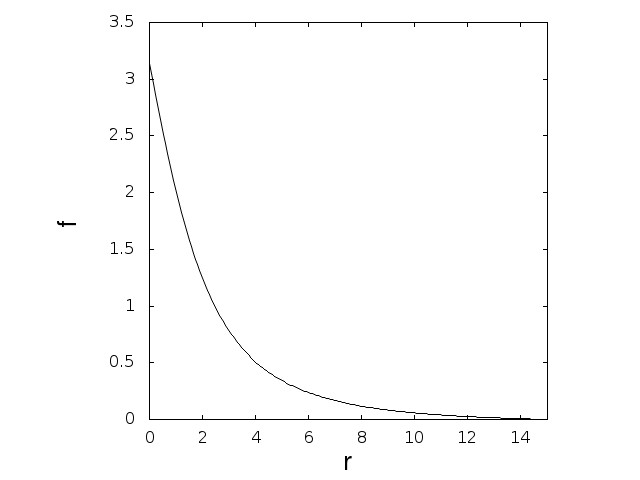}
\includegraphics[width=8cm]{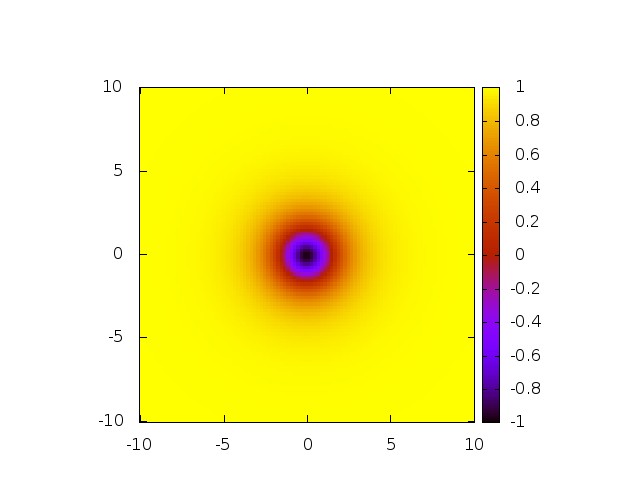}
}
\caption{The $N=1$ soliton. The left image is the profile function $f(r)$ and the right image is a plot of $\phi_3$ in the plane.}
\label{fig-B1}\end{center}\end{figure}

The radial profile function
satisfies a second order ordinary differential equation that is easily
derived by substituting the radial ansatz (\ref{radial}) into the energy
(\ref{energy}) and performing the variation. 
A numerical solution for the profile function with $N=1$ is shown in the
left image in Figure~\ref{fig-B1} and yields the 1-soliton energy $E_1=20.27.$
The right image in Figure~\ref{fig-B1} shows a plot of the function
$\phi_3$ in the plane, which is a convenient way to display the 
soliton solutions and will be used from now on. 

It turns out that radially symmetric solitons with $N>1$ are 
unstable in this theory and we must turn to numerical field theory
computations in the plane to study multi-solitons.
We apply a field theory energy minimization algorithm using
fourth order finite difference approximations for the spatial derivatives
on a lattice with $251\times 251$ grid points and a lattice spacing
$\Delta x=0.2.$ 
The field is fixed at the vacuum value $\bphi=(0,0,1)$ at the boundary of the grid.  
Note that, for clarity, the full extent of the simulation
region is not displayed in the figures that follow. 
The energy minimization algorithm is an
adaptation of the one described in detail in \cite{BS3}, 
where it is applied to the standard three-dimensional Skyrme model.
A range of charge $N$ initial conditions have been used, 
including perturbations that break the radial symmetry of configurations 
of the form (\ref{radial}), and $N$ well-separated single solitons with
random initial positions and phases.
\begin{figure}[ht]\begin{center}
\hbox{
\includegraphics[width=5.5cm]{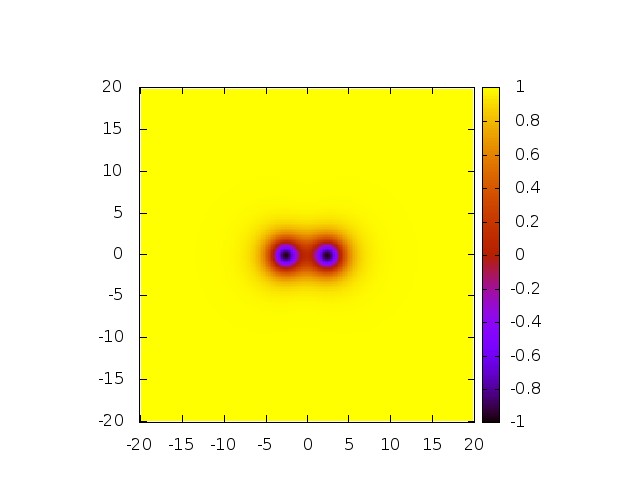}
\includegraphics[width=5.5cm]{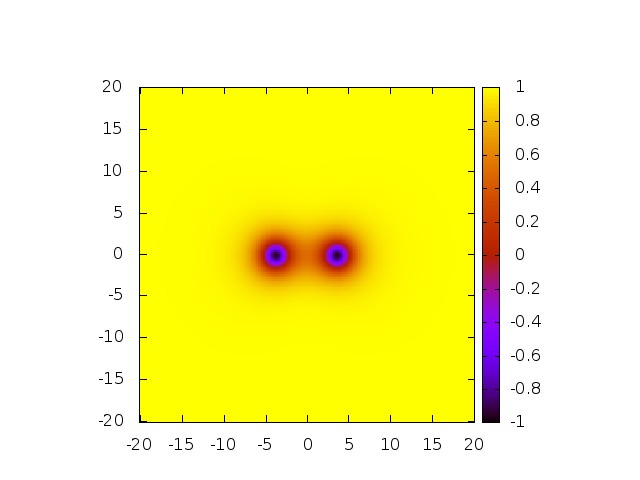}
\includegraphics[width=5.5cm]{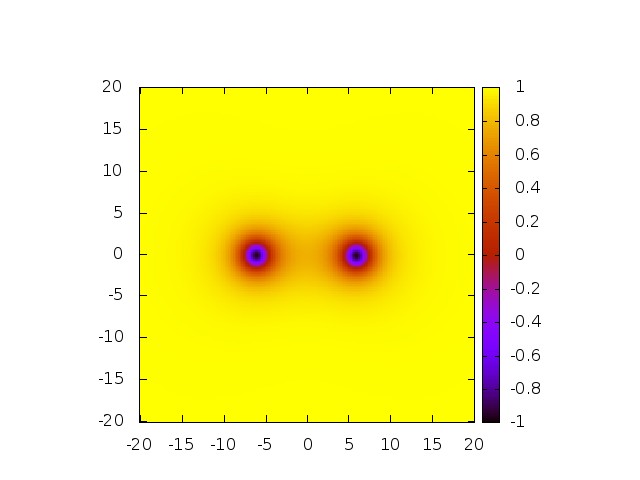}
}
\caption{A plot of $\phi_3$ for the 2-soliton
with $\mu=0.5,\,0.2,\,0.05$ (left to right).
}
\label{fig-mu}\end{center}\end{figure}

The minimal energy 2-soliton is 
displayed in the first image of Figure \ref{fig-mu}, 
where it can be seen that it resembles two separated single solitons. 
This 2-soliton has energy $E_2=39.84$, which is less than $2E_1,$ 
confirming that it is indeed a
bound state. Rather than list the energy $E$ of an $N$-soliton solution it is
more convenient to present the binding energy $\Delta=NE_1-E,$ which is
the energy required to split the bound state into $N$ single solitons.
The 2-soliton has binding energy $\Delta=2E_1-E_2=0.70,$
which is less than $2\%$ of the total energy, 
so the solitons are indeed weakly bound in this theory. 
 Although the two solitons form a bound state they 
keep their individual identities, remaining aloof rather than merging into
a radial configuration 
(which is the situation in the standard baby Skyrme model).
For comparison, the energy of the unstable
radially symmetric $N=2$ configuration is $E=40.49$, which is very  close to
twice the energy of a single soliton. 

As mentioned above, fixing the interpolation parameter value to $\mu=0.5$ is
sufficient to induce the features that we wish to highlight. However, to
demonstrate that there is nothing special about this choice we display the
2-soliton solution for the parameter values $\mu=0.2$ and $\mu=0.05.$ as
the second and third images in Figure \ref{fig-mu}. As expected, as $\mu$
decreases towards the repulsive limit $\mu=0$ the solitons becomes 
increasingly aloof, with the separation between the constituent solitons
increasing and binding energies decreasing. From now on we shall consider
only the value $\mu=0.5,$ being aware that the qualitative features should
persist for all sufficiently small values of $\mu$.

The static 2-soliton solution suggests that there is a short-range
repulsion and a long-range attraction between two single solitons.
The long-range attraction can be understood analytically by calculating
the asymptotic interaction energy between two well-separated solitons at
the linear level. This calculation has been performed for the standard
baby Skyrme model \cite{PSZ} and the result can be applied directly to the
aloof model with $\mu\in(0,1]$, because the  
last term in the potential (\ref{pot3}) does not contribute to
the interaction potential at the linear level.
For two solitons with a separation $R$ and a relative phase $\chi$ the 
asymptotic result for the large $R$ interaction energy 
$U^{\rm asy}\sim E-2E_1$ is \cite{PSZ}
\be
U^{\rm asy}(R,\chi)=C\cos\chi\frac{e^{-m\sqrt{\mu}R}}{\sqrt{R}},
\ee
where $C$ is a positive constant that depends on $m$ and $\mu.$
\begin{figure}[ht]\begin{center}
\includegraphics[width=8cm]{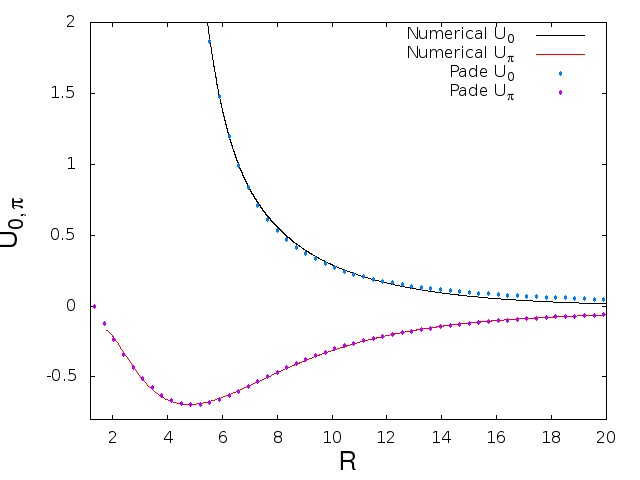}
\caption{The numerically computed interaction potential 
$U_0$ for two solitons that are in phase (upper blue curve) and 
the interaction potential $U_\pi$ for two solitons that are 
out of phase (lower red curve). 
Dots denote the Pad\'e approximants of order [3/4] described in the text. 
}
\label{fig-potentials}\end{center}\end{figure}

This formula shows that two solitons that are in phase ($\chi=0$) are
repulsive, as the interaction energy is positive. The most attractive channel
is when the two solitons are exactly out of phase ($\chi=\pi$), so that
the interaction energy is maximally negative and there is an
attractive force at large separations. 
The fact that this long-range attractive force turns into a short-range
repulsive force cannot be determined by a linear analysis because it is
due to the last term in the potential (\ref{pot3}), 
which is a nonlinear effect as the linear contribution vanishes \cite{PSZ}. 
We therefore turn to a numerical computation of the interaction potential.

First we compute the interaction potential, $U_0(R)$, for two solitons with separation $R$ and relative phase $\chi=0$. 
An initial condition containing two solitons with
given positions and phases can be constructed using a product ansatz as
follows. Introduce the Riemann sphere coordinate $W=(\phi_1+i\phi_2)/(1-\phi_3)$, obtained by stereographic projection of $\bphi.$ 
Let $W_1$ and $W_2$ be the associated fields of two $N=1$ solitons, with
arbitrary positions and phases, and construct the $N=2$ field $W$ via
\be
W=\frac{W_1W_2}{W_1+W_2},
\ee
which provides $\bphi$ by inverse stereographic projection. We take  
this form of initial condition with relative phase $\chi=0$ and a range
of small initial separations $R_0\in[3,5],$ and apply our energy minimization
algorithm, computing the separation $R$ and energy $E$ as the minimization
takes place. As there is a repulsive force between the solitons, $R$ increases
and $E$ decreases, producing the in phase interaction potential
$U_0(R)=E-2E_1$ displayed as the upper black curve in Figure 
\ref{fig-potentials}. The computed potential $U_0(R)$ is independent of the
initial separation $R_0$ over the range shown, where $R>R_0.$ 

To compute the interaction potential, $U_\pi(R)$, for two solitons with separation $R$ and relative phase $\chi=\pi$, we use the same product ansatz 
to compute an initial condition for two out of phase solitons, but this time
with a large initial separation $R_0.$ The results presented used the value
 $R_0=20,$ but the results are independent of $R_0$ providing it is sufficiently
large. This time, rather than using an energy minimization algorithm, we 
evolve the full nonlinear, energy conserving, second order time dependent
field equations that follow from the relativistic baby Skyrmion 
Lagrangian associated with the static energy (\ref{energy}). 
As the inter-soliton force is attractive at long-range, the soliton separation $R$ decreases with time. Although the total energy is conserved in relativistic
dynamics (and is conserved to a high accuracy in our numerical computations)
the static energy $E$ given by (\ref{energy}) is not conserved 
because of the missing kinetic energy contribution. Monitoring the separation
$R$ and the static energy $E$ as a function of time results in the plot
of the interaction potential $U_\pi(R)=E-2E_1,$ displayed as the 
lower red curve in Figure \ref{fig-potentials}.      

We have been able to compute the interaction potential using this method by
exploiting the fact that baby Skyrmion dynamics produces very little
radiation, and binding energies are small hence the soliton speeds remain low.  
We see from the curve for $U_\pi(R)$ in Figure \ref{fig-potentials} that the
turning point is reproduced at the correct soliton separation and binding
energy of the static 2-soliton solution. Furthermore, the short-range
repulsion is now evident, and the reason we are able to compute the 
potential $U_\pi(R)$ almost all the way to $R=0$ is because the 
unstable radially symmetric $N=2$ configuration has an energy very close
to twice that of the 1-soliton.       

The numerical computations confirm that solitons that are in phase
$(\chi=0)$ are repulsive and solitons that are out of phase ($\chi=\pi$)
have a long-range attraction and a short-range repulsion. In what follows it
will be useful to have an analytic approximation to the interaction
potentials $U_\chi(R)$ for these two relative phases $\chi=0,\pi.$
In the range of $R$ of interest 
(essentially the range displayed in Figure \ref{fig-potentials}), 
both interaction potentials can be described
by a Pad\'e approximant of order [3/4] given by
\be
U_{\chi}(R)=\frac{\sum_{j=0}^3 a_jR^j}{\sum_{j=0}^4 b_jR^j}. 
\label{pade}
\ee
For $\chi=\pi$ we make use of the fact that the energy of the $N=2$ radial
solution is approximately twice the 1-soliton energy to impose the condition
$U_\pi(0)=0,$ which fixes the constant $a_0=0,$ and without loss of 
generality we set the normalisation by choosing $b_0=1.$
For $\chi=0$ we require that $U_0(R)$ diverges as $R\to 0,$ hence we fix
$b_0=0$ and set the normalisation by choosing $a_0=1.$
The remaining constants can be found in Table \ref{tab-pade} and were 
determined by a least squares fit to the numerical data presented 
in Figure \ref{fig-potentials}. The dots in Figure \ref{fig-potentials} 
show the results of evaluating these Pad\'e approximants and confirm that
they provide good approximations to the interaction potentials. 
\begin{table}[ht]
\centering
\begin{tabular}{|c|c|c|c|c|c|c|c|c|c|}\hline
& $a_0$ & $a_1$ & $a_2$ & $a_3$ & $b_0$ &  $b_1$ &  $b_2$ &  $b_3$ &  $b_4$\\ \hline 
$U_{\pi}$ 
& 0 & 1.355 & -1.024 & 0.012 & 1 & 0.716 & 1.271 & -0.298 & 0.041 \\ \hline
$U_{0}$
& 1 & 8.469 & 3.904 & -0.019 & 0 &
2.583 & -0.609 & -0.868 & 0.254
 \\ \hline
\end{tabular}
\caption{The coefficients in the [3/4] Pad\'e approximants of the interaction
potentials $U_{\pi}, U_{0}$.}
 \label{tab-pade}
\end{table}

\begin{figure}
\centering
\begin{subfigure}[b]{0.238\textwidth}\includegraphics[width=\textwidth]{B2.jpg}\caption{2}\label{B2}\end{subfigure}
\begin{subfigure}[b]{0.244\textwidth}\includegraphics[width=\textwidth]{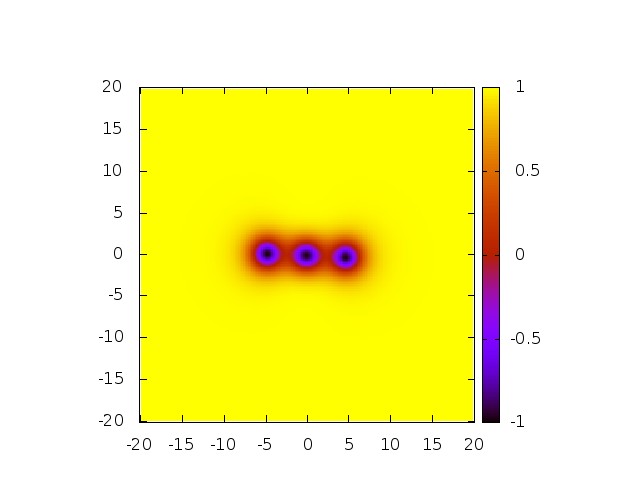}\caption{3}\end{subfigure}
\begin{subfigure}[b]{0.244\textwidth}\includegraphics[width=\textwidth]{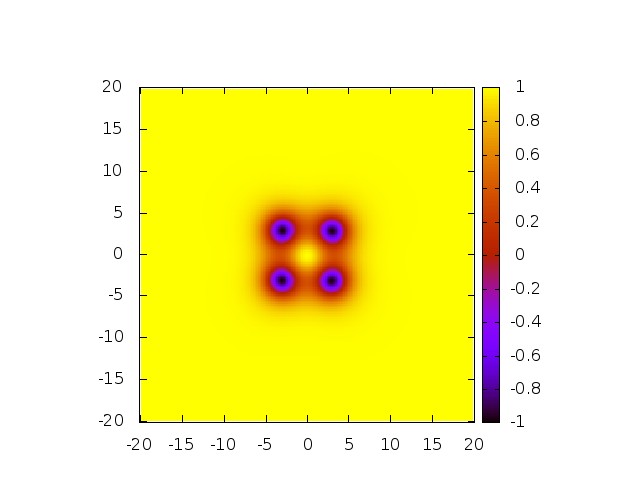}\caption{4}\end{subfigure}
\begin{subfigure}[b]{0.244\textwidth}\includegraphics[width=\textwidth]{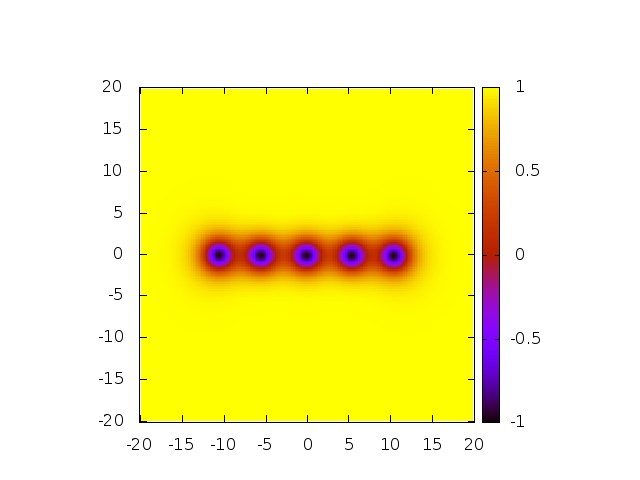}\caption{5}\end{subfigure}
\begin{subfigure}[b]{0.244\textwidth}\includegraphics[width=\textwidth]{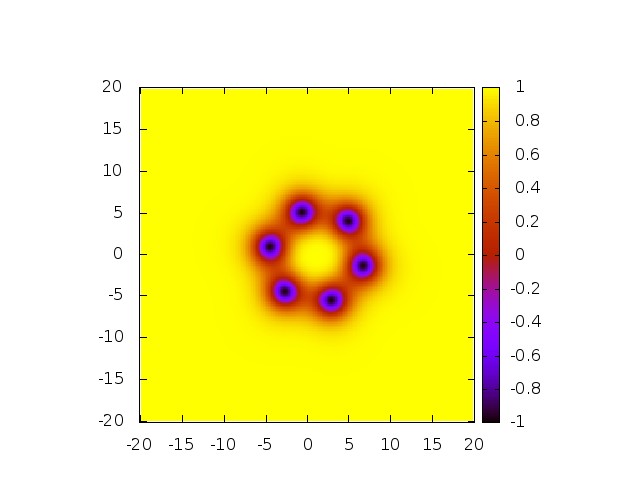}\caption{6}\end{subfigure}
\begin{subfigure}[b]{0.238\textwidth}\includegraphics[width=\textwidth]{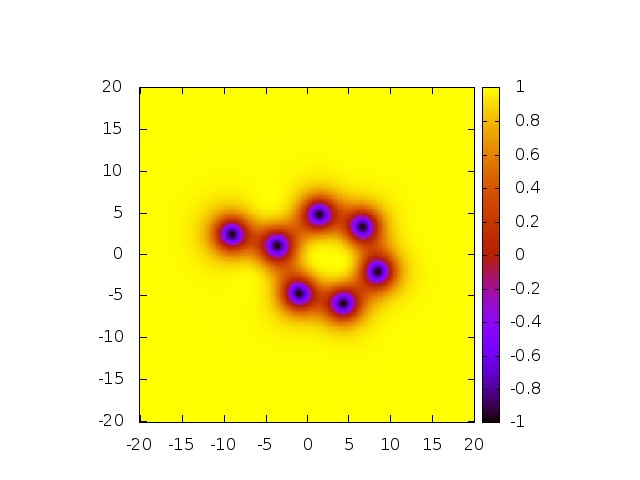}\caption{7}\end{subfigure}
\begin{subfigure}[b]{0.23\textwidth}\includegraphics[width=\textwidth]{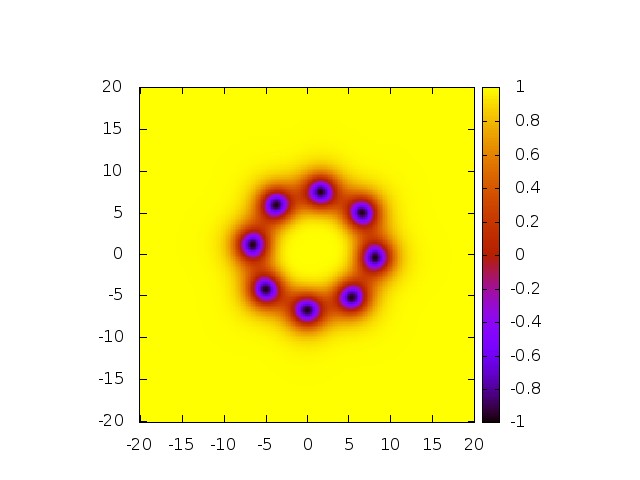}\caption{8}\end{subfigure}
\begin{subfigure}[b]{0.23\textwidth}\includegraphics[width=\textwidth]{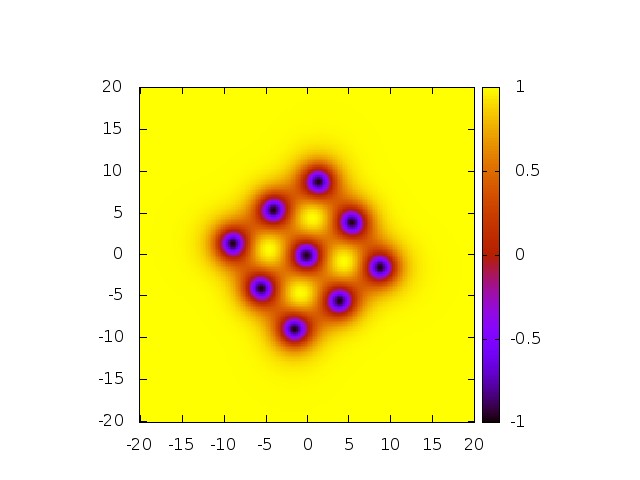}\caption{9}\end{subfigure}
\begin{subfigure}[b]{0.23\textwidth}\includegraphics[width=\textwidth]{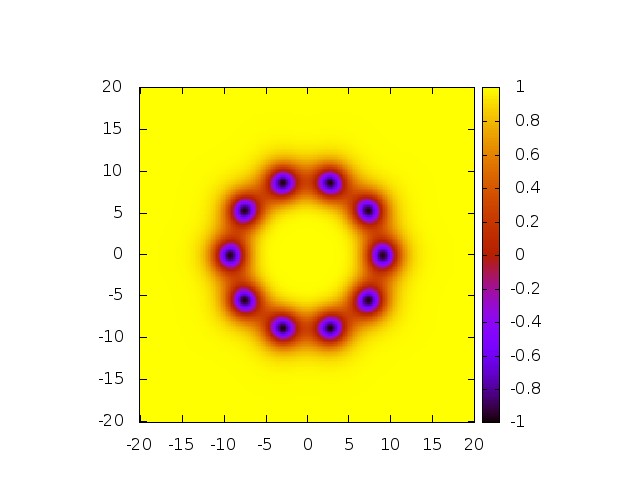}\caption{10}\end{subfigure}
\begin{subfigure}[b]{0.23\textwidth}\includegraphics[width=\textwidth]{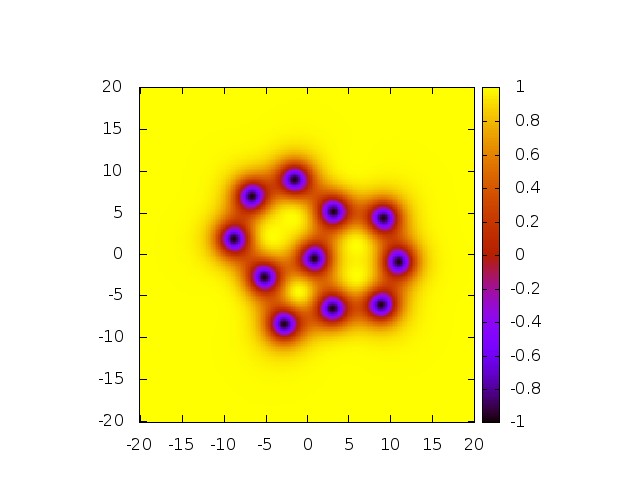}\caption{11}\end{subfigure}
\begin{subfigure}[b]{0.23\textwidth}\includegraphics[width=\textwidth]{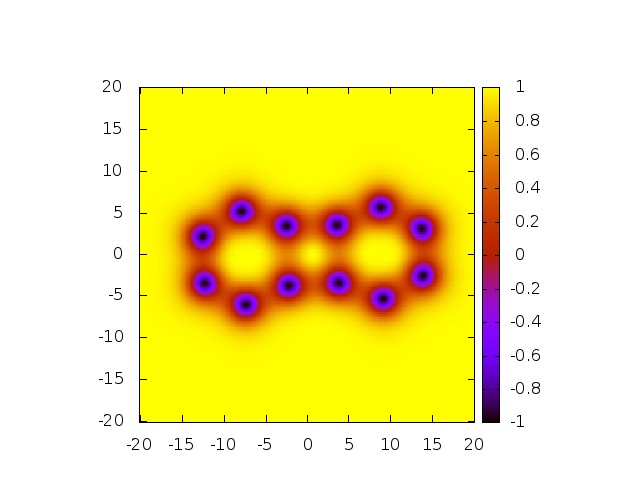}\caption{12}\end{subfigure}
        \caption{Plots of $\phi_3$ for global energy minima for $N$-solitons with $2\le N\le 12$.}
\label{fig-solitons}
\end{figure}

Turning now to solitons of higher charge, in Figure \ref{fig-solitons} we 
present the global minimal energy static $N$-soliton for $2\le N \le 12.$
Most of these solutions were obtained by applying our field theory energy
minimization algorithm to initial conditions containing $N$ single solitons
with random positions and internal phases. Many local energy minima were also
obtained and we shall discuss this aspect further in the following section,
together with more details regarding the construction of initial conditions.

The global minima displayed in Figure \ref{fig-solitons} reveal a range of
configuration types including linear chains, circular necklaces, square
arrays and cluster structures. The 7-soliton is an interesting example 
and resembles a soliton analogue of a halo nucleus, 
with 6 solitons in the core and a weakly bound single soliton in the halo.
The 12-soliton clearly has a cluster decomposition into two 6-solitons, 
being reminiscent of the recurrent clustering phenomenon in light nuclei.
The fact that the basic building block in our baby Skyrme model 
appears to be the 6-soliton, compared to the 4-nucleon alpha-particle 
in nuclei, is likely a result of the two-dimensional nature of our toy model.
\begin{figure}[ht]\centering
\includegraphics[width=8cm]{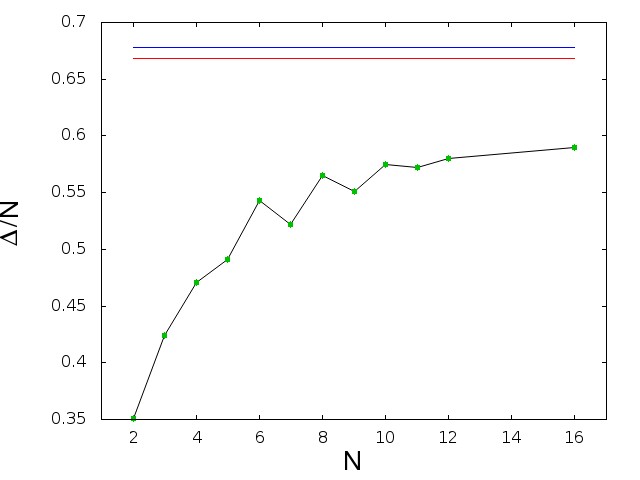}
\caption{The binding energy per soliton $\Delta/N$. 
The upper blue and red lines denote the values for the hexagonal and
square lattice respectively.}
\label{fig-bind}\end{figure}

In Figure \ref{fig-bind} we plot the binding energy per soliton $\Delta/N$
for these minimal energy $N$-soliton solutions (the precise values for
$\Delta$ will be presented in a later table). 
The plot displays the correct qualitative features 
of nuclear binding energies (in the absence of Coulomb effects) 
with a general increase of the binding energy per nucleon 
towards a plateau, and tighter binding for even numbers of nucleons
in comparison to neighbouring odd numbers.   

The variety of pattern types displayed in Figure \ref{fig-solitons} make
it far from obvious to predict the form taken by the minimal 
energy $N$-soliton for any given value of $N.$ However, the fact that
the constituent single solitons remain aloof 
means that a point particle model may be able
to predict the structures obtained from the field theory computations.
In the following section we introduce a binary species point particle
model and demonstrate that it indeed provides an excellent approximation
and is able to predict the form of all the soliton solutions obtained.

\section{A binary species particle model}\quad
As we have seen, the attractive channel for two solitons is when the relative
phase between the two solitons is equal to $\pi.$ This suggests that 
the constituent single solitons in an $N$-soliton solution can be allocated
into two groups, where all solitons in a given group have the same phase
and there is a relative phase of $\pi$ between two solitons in different groups.
An appropriate arrangement of the solitons will then allow a large number of 
attractive channel pairings, with repulsive pairings being suppressed by
optimal spatial positioning. 

Our point particle model makes this assumption and is a binary
species model where we label the two species as blue and red. 
Blue particles represent single solitons with an internal phase $\chi=0$
and red particles are solitons with an internal phase $\chi=\pi.$
Two particles of different colours are therefore in the attractive
channel and we define the potential between them to be given by the
Pad\'e approximant (\ref{pade}) to the potential $U_\pi(R),$ where
$R$ is the distance between the two point particles. Similarly, we 
define the potential between two particles of the same colour to be
the Pad\'e approximant to the potential $U_0(R)$.
To create as many attractive pairs as possible we alternately label
the particles as blue and red, so that there are $N_{\rm red}=\lfloor{N/2}\rfloor$
red particles and $N_{\rm blue}=N-N_{\rm red}$ blue particles.
The particle binding energy $\delta$ is simply defined to be minus 
the sum over all pairs of interactions, that is
\be
\delta=-\sum_{i=2}^N\sum_{j=1}^{i-1}U_{\chi_{ij}}(|{\bf x}^{(i)}-{\bf x}^{(j)}|),
\ee
where ${\bf x}^{(1)},\ldots,{\bf x}^{(N)}$ are the positions of the $N$ particles
in the plane and $\chi_{ij}=0$ if particles with positions 
${\bf x}^{(i)}$ and ${\bf x}^{(j)}$ are of the same colour and 
$\chi=\pi$ otherwise.

\renewcommand{\thesubfigure}{\roman{subfigure}}

\begin{figure}
\centering
\begin{subfigure}[b]{0.17\textwidth}\includegraphics[width=\textwidth]{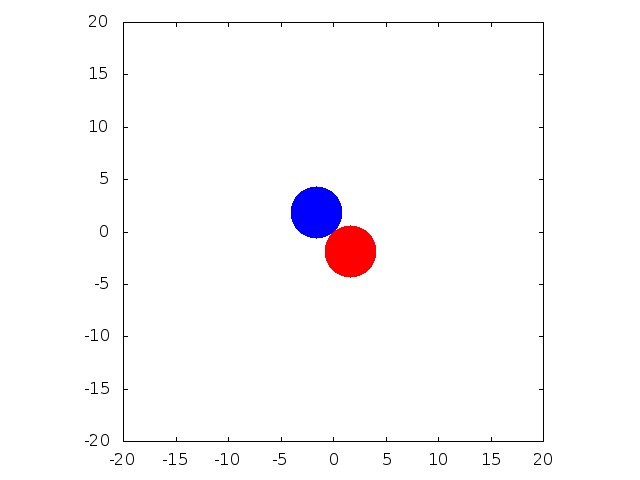}\caption{2*}\end{subfigure}
\begin{subfigure}[b]{0.17\textwidth}\includegraphics[width=\textwidth]{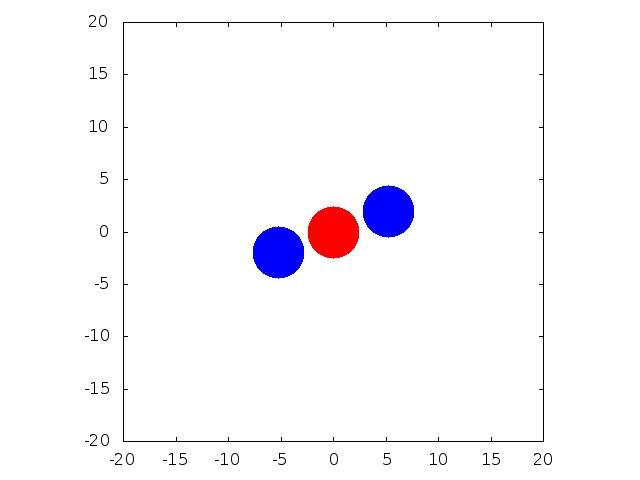}\caption{3*}\end{subfigure}
\begin{subfigure}[b]{0.17\textwidth}\includegraphics[width=\textwidth]{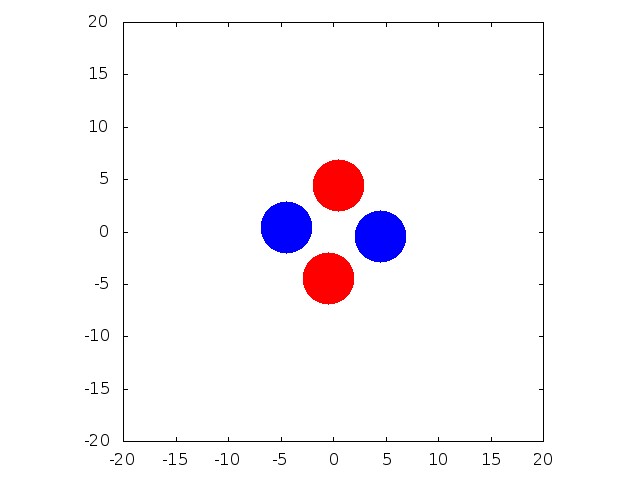}\caption{4*}\end{subfigure}
\begin{subfigure}[b]{0.17\textwidth}\includegraphics[width=\textwidth]{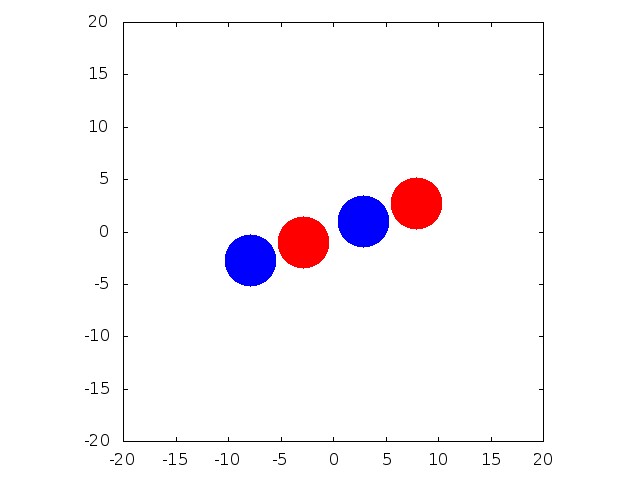}\caption{4}\end{subfigure}
\begin{subfigure}[b]{0.17\textwidth}\includegraphics[width=\textwidth]{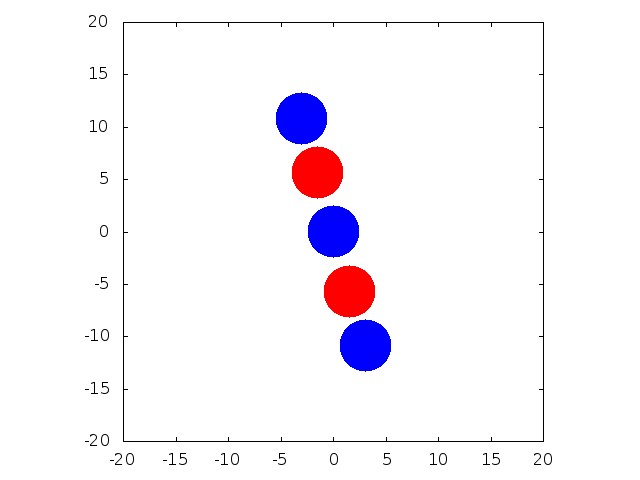}\caption{5*}\end{subfigure}
\begin{subfigure}[b]{0.17\textwidth}\includegraphics[width=\textwidth]{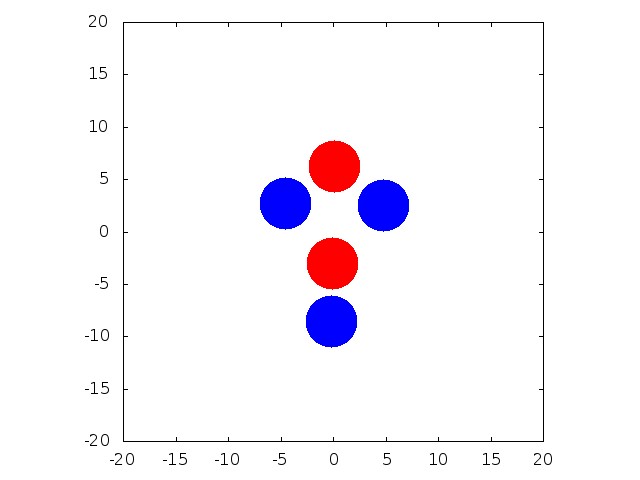}\caption{5}\end{subfigure}
\begin{subfigure}[b]{0.17\textwidth}\includegraphics[width=\textwidth]{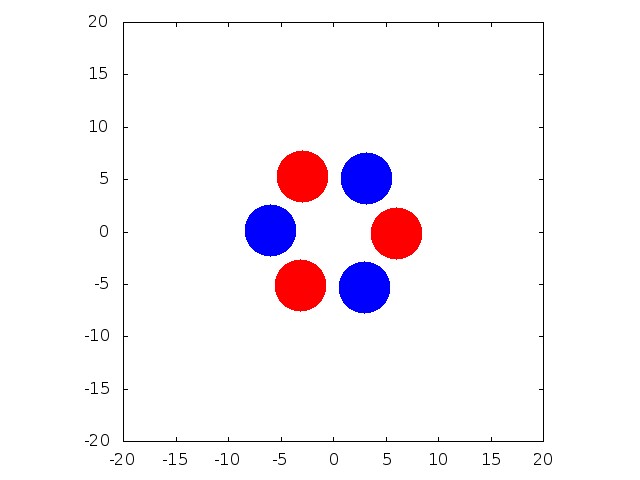}\caption{6*}\end{subfigure}
\begin{subfigure}[b]{0.17\textwidth}\includegraphics[width=\textwidth]{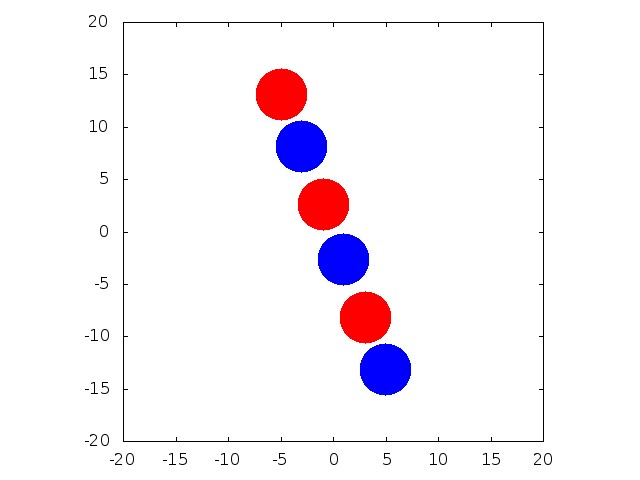}\caption{6}\end{subfigure}
\begin{subfigure}[b]{0.17\textwidth}\includegraphics[width=\textwidth]{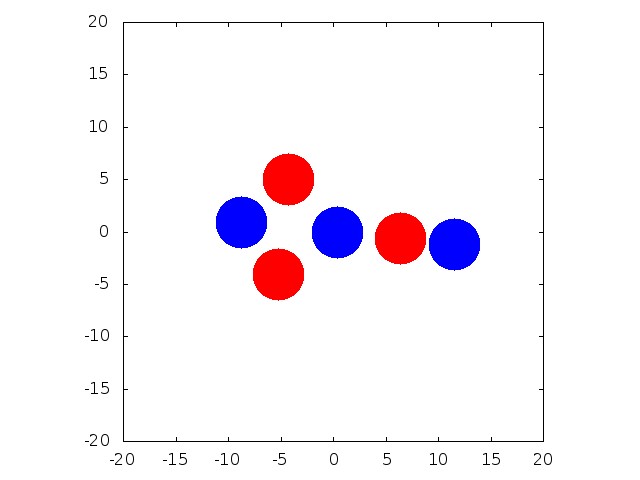}\caption{6}\end{subfigure}
\begin{subfigure}[b]{0.17\textwidth}\includegraphics[width=\textwidth]{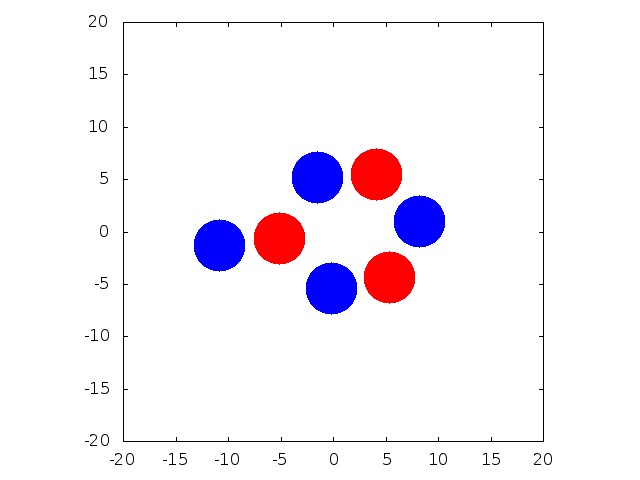}\caption{7*}\end{subfigure}
\begin{subfigure}[b]{0.17\textwidth}\includegraphics[width=\textwidth]{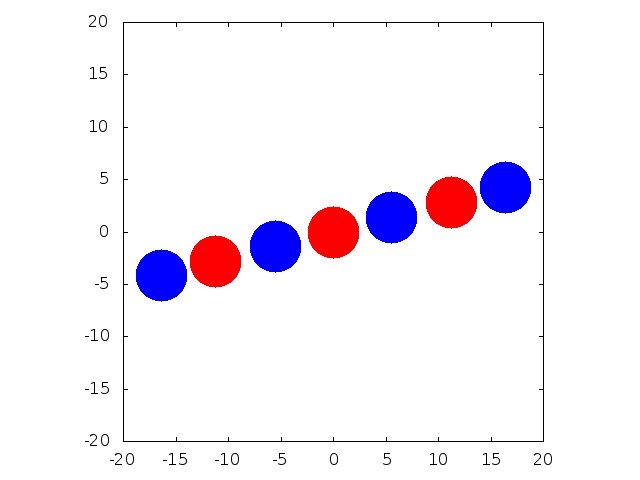}\caption{7}\end{subfigure}
\begin{subfigure}[b]{0.17\textwidth}\includegraphics[width=\textwidth]{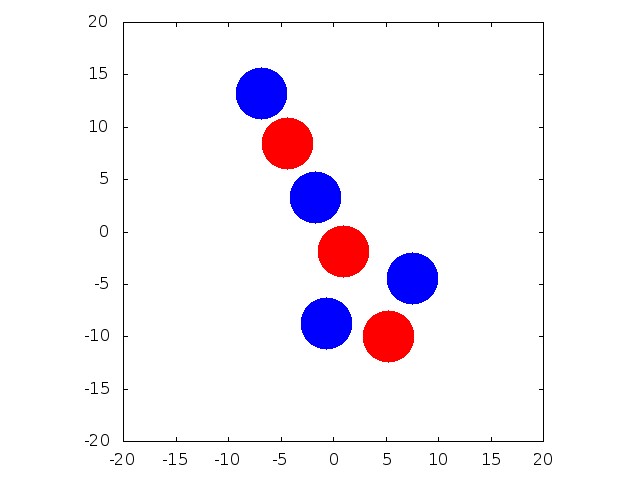}\caption{7}\end{subfigure}
\begin{subfigure}[b]{0.17\textwidth}\includegraphics[width=\textwidth]{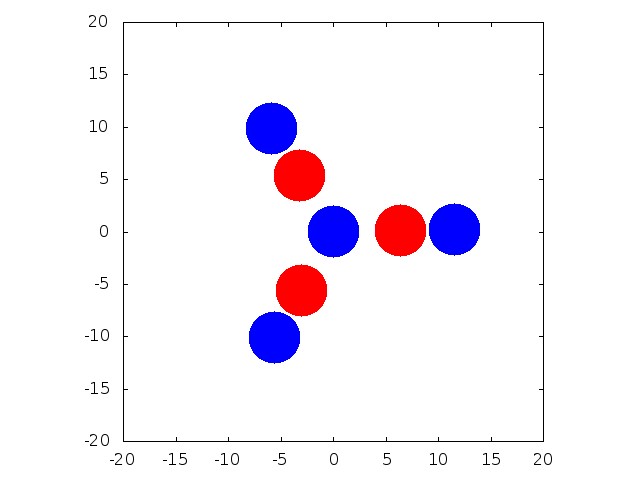}\caption{7}\end{subfigure}
\begin{subfigure}[b]{0.17\textwidth}\includegraphics[width=\textwidth]{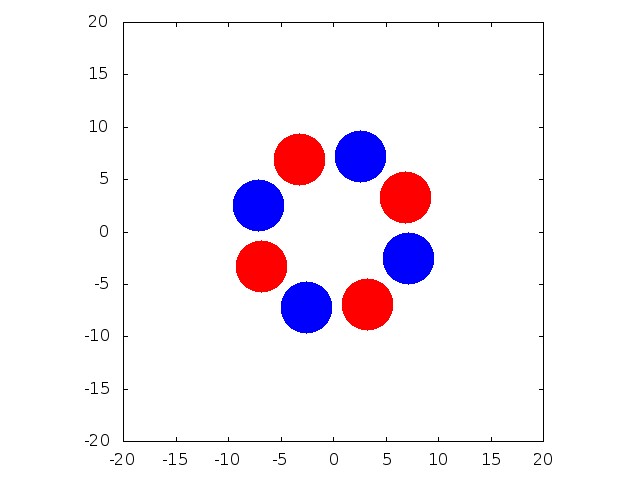}\caption{8*}\end{subfigure}
\begin{subfigure}[b]{0.17\textwidth}\includegraphics[width=\textwidth]{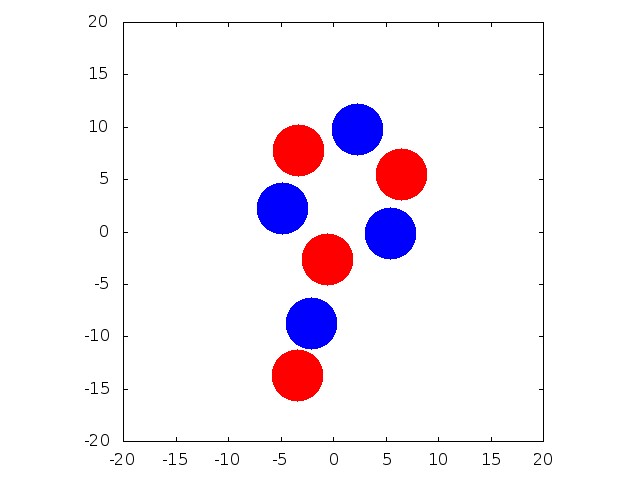}\caption{8}\end{subfigure}
\begin{subfigure}[b]{0.17\textwidth}\includegraphics[width=\textwidth]{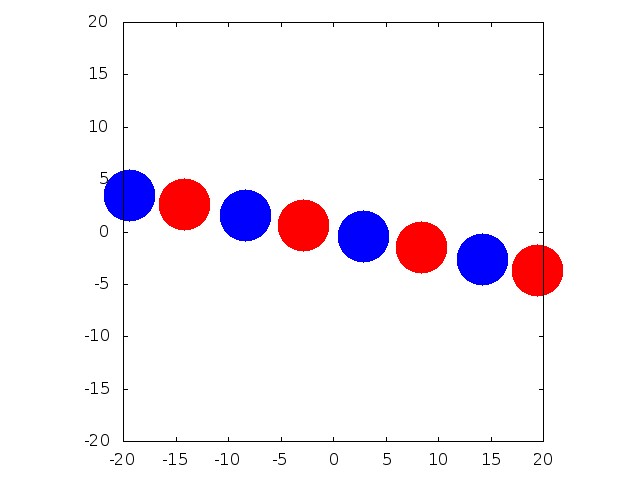}\caption{8}\end{subfigure}
\begin{subfigure}[b]{0.17\textwidth}\includegraphics[width=\textwidth]{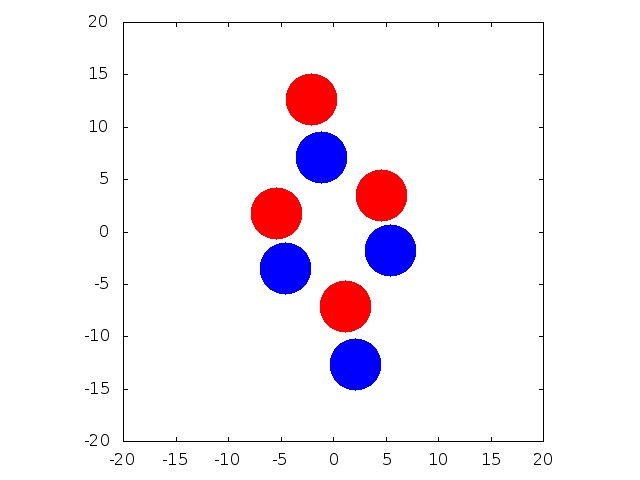}\caption{8}\end{subfigure}
\begin{subfigure}[b]{0.17\textwidth}\includegraphics[width=\textwidth]{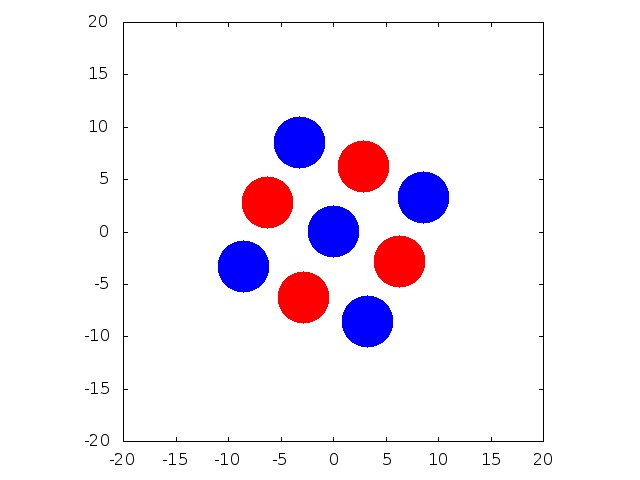}\caption{9*}\end{subfigure}
\begin{subfigure}[b]{0.17\textwidth}\includegraphics[width=\textwidth]{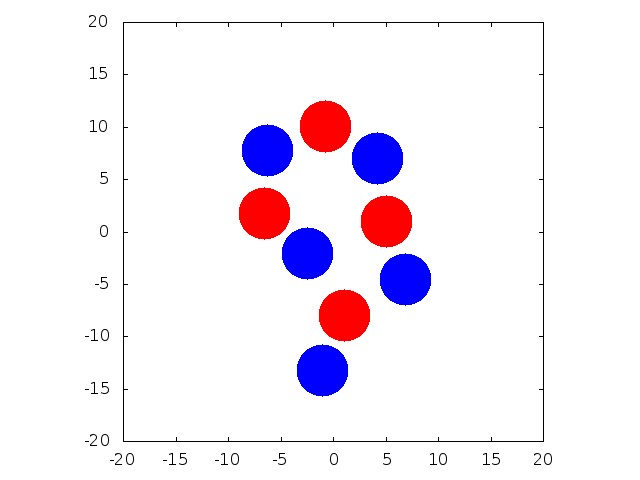}\caption{9}\end{subfigure}
\begin{subfigure}[b]{0.17\textwidth}\includegraphics[width=\textwidth]{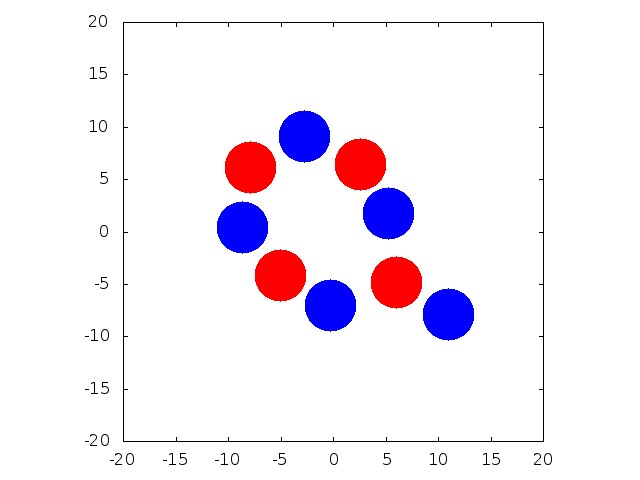}\caption{9}\end{subfigure}
\begin{subfigure}[b]{0.17\textwidth}\includegraphics[width=\textwidth]{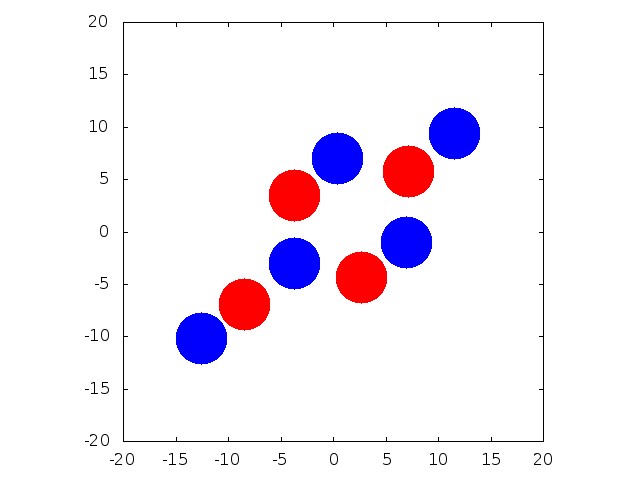}\caption{9}\end{subfigure}
\begin{subfigure}[b]{0.17\textwidth}\includegraphics[width=\textwidth]{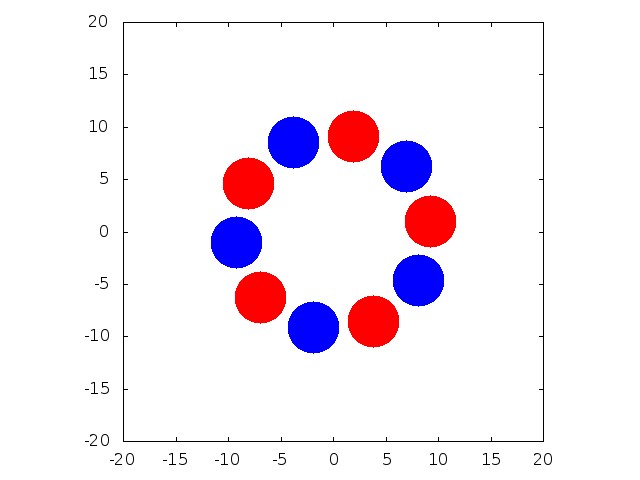}\caption{10*}\end{subfigure}
\begin{subfigure}[b]{0.17\textwidth}\includegraphics[width=\textwidth]{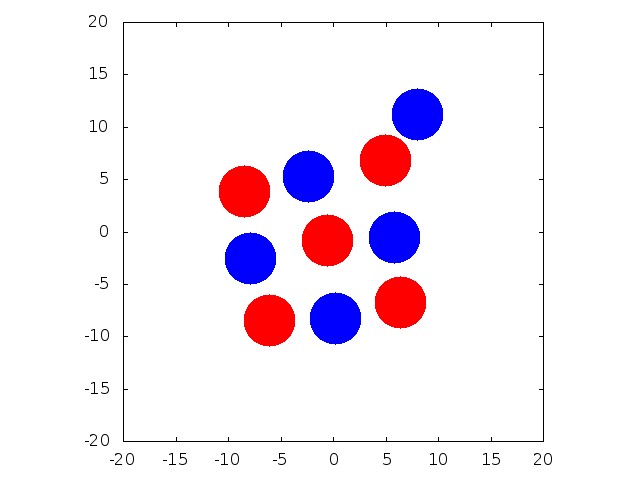}\caption{10}\end{subfigure}
\begin{subfigure}[b]{0.17\textwidth}\includegraphics[width=\textwidth]{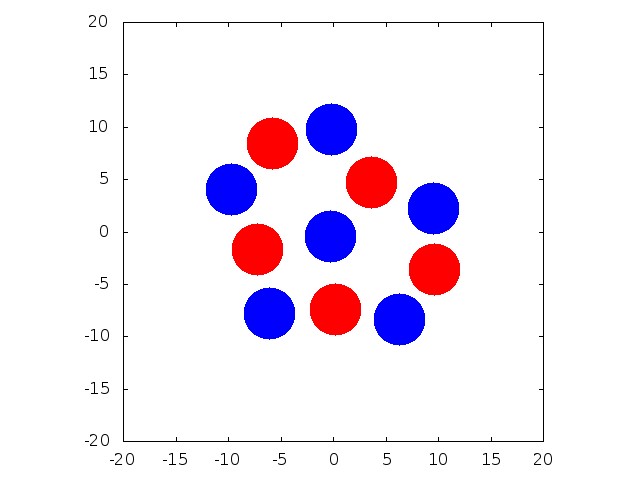}\caption{11*}\end{subfigure}
\begin{subfigure}[b]{0.17\textwidth}\includegraphics[width=\textwidth]{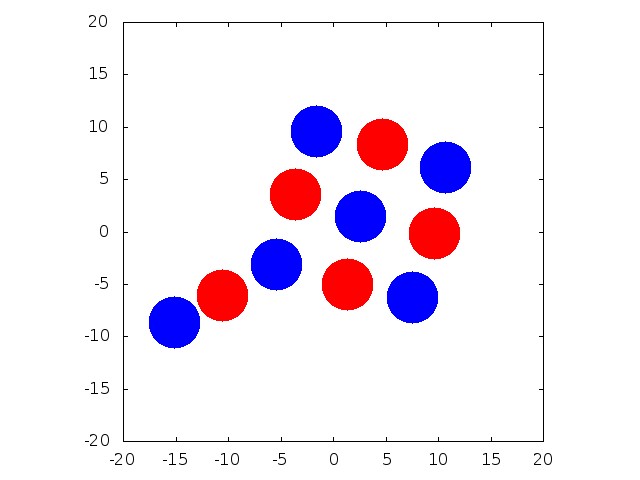}\caption{11}\end{subfigure}
\begin{subfigure}[b]{0.17\textwidth}\includegraphics[width=\textwidth]{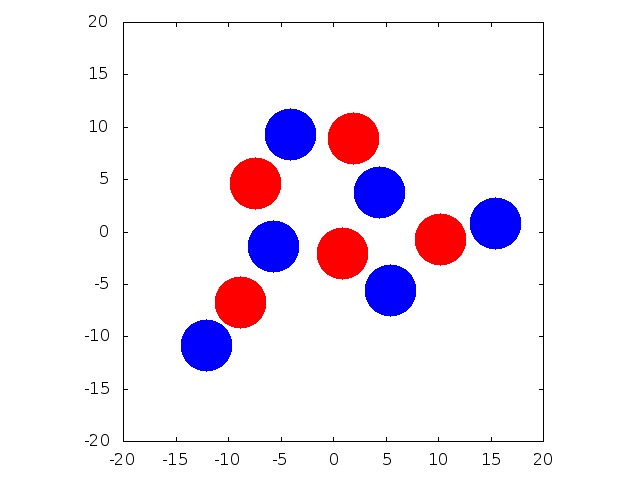}\caption{11}\end{subfigure}
\begin{subfigure}[b]{0.17\textwidth}\includegraphics[width=\textwidth]{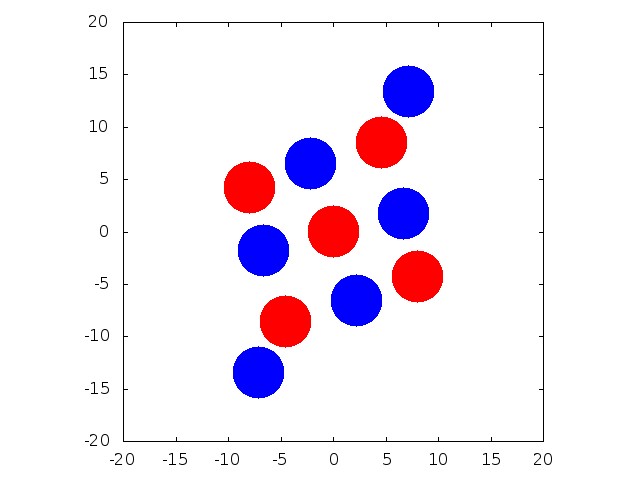}\caption{11}\end{subfigure}
\begin{subfigure}[b]{0.17\textwidth}\includegraphics[width=\textwidth]{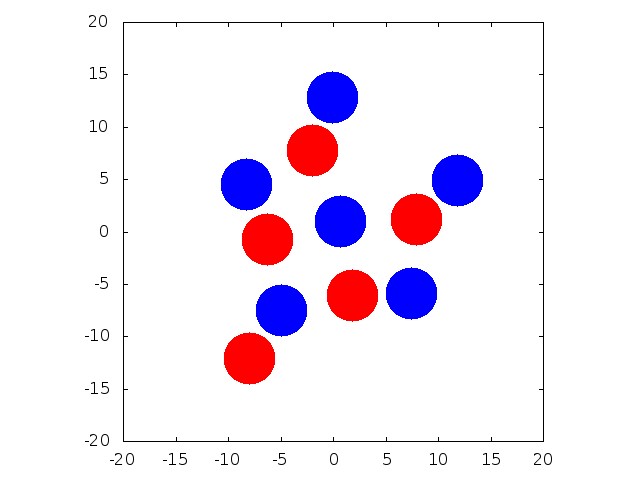}\caption{11}\end{subfigure}
\begin{subfigure}[b]{0.17\textwidth}\includegraphics[width=\textwidth]{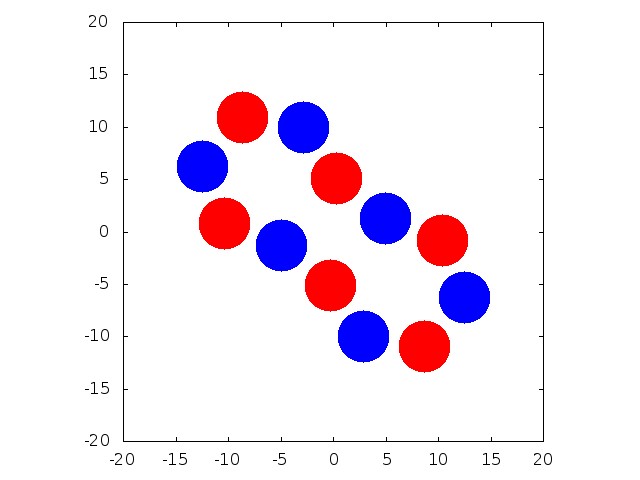}\caption{12*}\end{subfigure}
\begin{subfigure}[b]{0.17\textwidth}\includegraphics[width=\textwidth]{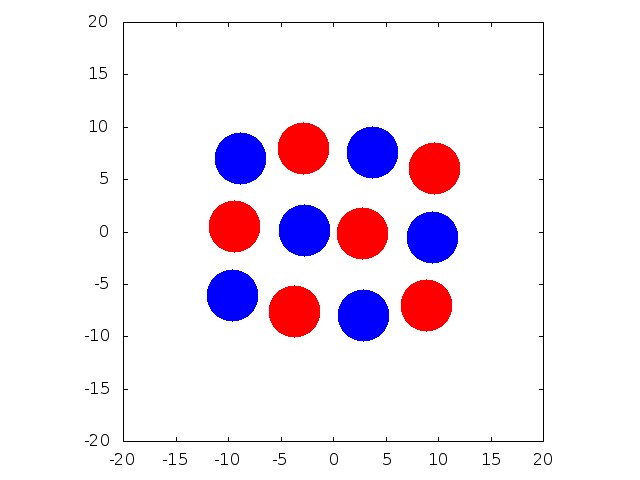}\caption{12}\end{subfigure}
\begin{subfigure}[b]{0.17\textwidth}\includegraphics[width=\textwidth]{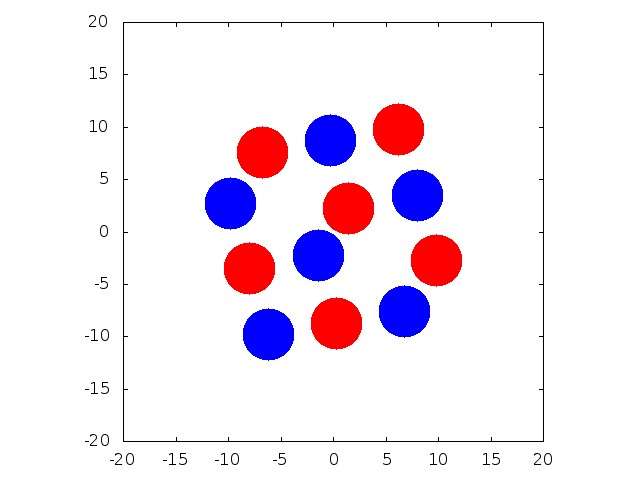}\caption{12}\end{subfigure}
\begin{subfigure}[b]{0.17\textwidth}\includegraphics[width=\textwidth]{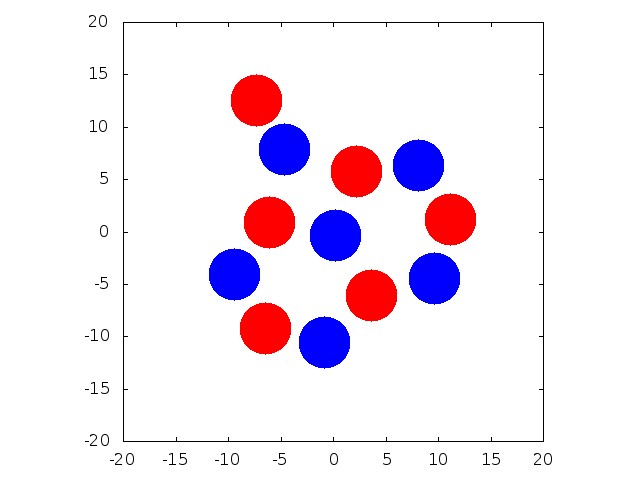}\caption{12}\end{subfigure}
\begin{subfigure}[b]{0.17\textwidth}\includegraphics[width=\textwidth]{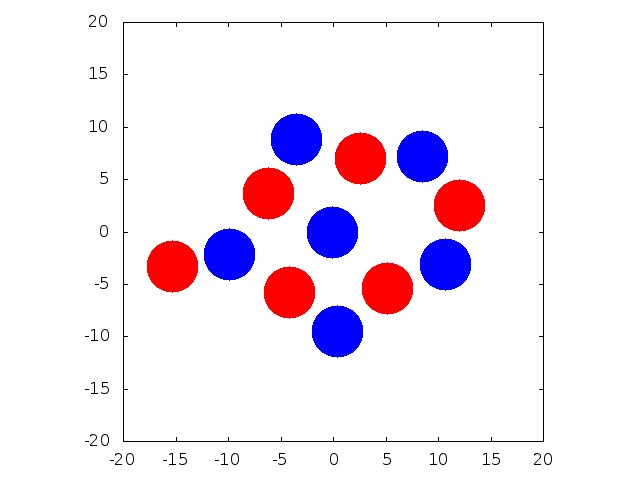}\caption{12}\end{subfigure}
\begin{subfigure}[b]{0.17\textwidth}\includegraphics[width=\textwidth]{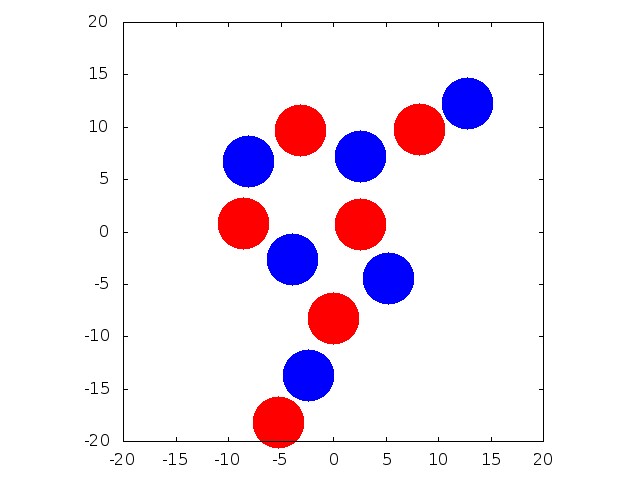}\caption{12}\end{subfigure}
\ \hskip 2.5cm \
        \caption{Local maxima of the binding energy $\delta$ in the $N$-particle model with $2\le N\le 12$. \\
* denotes the global maximum of $\delta$ for each value of $N$.}
\label{fig-particles}
\end{figure}

\begin{table}[ht]
\centering
\begin{tabular}{|r|c|c|r|}\hline
Configuration & $\Delta$ & $\delta$ & \% \\ \hline
(i) \ 2* \includegraphics[width=0.58cm]{b2a.jpg} &    0.701 & 0.698 & 100.0 \\ \hline
(ii) \ 3* \includegraphics[width=0.58cm]{b3a.jpg} &    1.272 & 1.153 & 100.0 \\ \hline
(iii) \ 4* \includegraphics[width=0.58cm]{b4a.jpg}&    1.884 & 1.740 & 21.7 \\ \hline
(iv) \ 4\ \, \includegraphics[width=0.58cm]{b4b.jpg}& 1.868 & 1.733 & 78.3 \\ \hline
(v) \ 5* \includegraphics[width=0.58cm]{b5a.jpg} &   2.457 & 2.260 & 93.9\\ \hline
(vi) \ 5\ \, \includegraphics[width=0.58cm]{b5b.jpg}& 2.367 & 2.174 & 6.1\\ \hline
(vii) \ 6* \includegraphics[width=0.58cm]{b6a.jpg} &   3.259 & 3.036 & 79.0\\ \hline
(viii) \ 6\ \, \includegraphics[width=0.58cm]{b6b.jpg}& 3.042 & 2.820 & 11.2\\ \hline
(ix) \ 6\ \, \includegraphics[width=0.58cm]{b6c.jpg}& 2.982     & 2.788 & 9.8\\ \hline
(x) \ 7* \includegraphics[width=0.58cm]{b7a.jpg} &   3.651 & 3.407 & 73.6 \\ \hline
(xi) \ 7\ \, \includegraphics[width=0.58cm]{b7b.jpg} & 3.620 & 3.362 & 19.6 \\ \hline
(xii) \ 7\ \, \includegraphics[width=0.58cm]{b7c.jpg} & 3.564 & 3.310 & 4.4\\ \hline
(xiii) \ 7\ \, \includegraphics[width=0.58cm]{b7d.jpg} & 3.406 & 3.264 & 2.4\\ \hline
(xiv) \ 8* \includegraphics[width=0.58cm]{b8a.jpg} & 4.522 & 4.278 & 93.1\\ \hline
(xv) \ 8\ \, \includegraphics[width=0.58cm]{b8b.jpg} & 4.281 & 4.060 & 1.0 \\ \hline
(xvi) \ 8\ \, \includegraphics[width=0.58cm]{b8c.jpg} & 4.100 & 3.916 & 0.4\\ \hline
(xvii) \ 8\ \, \includegraphics[width=0.58cm]{b8d.jpg} & 4.053 & 3.837 & 5.5\\ \hline
(xviii) \ 9* \includegraphics[width=0.58cm]{b9a.jpg} & 4.958 & 4.656 & 8.2 \\ \hline
(xix) \ 9\ \, \includegraphics[width=0.58cm]{b9b.jpg} & 4.906 & 4.632 & 82.0 \\ \hline
(xx) \ 9\ \, \includegraphics[width=0.58cm]{b9c.jpg} & 4.877 & 4.629 & 8.4 \\ \hline
(xxi) \ 9\ \, \includegraphics[width=0.58cm]{b9d.jpg} & 4.684 & 4.462 & 1.4 \\ \hline
(xxii) 10* \includegraphics[width=0.58cm]{b10a.jpg} & 5.750 & 5.466 & 85.2\\ \hline
(xxiii) 10\ \, \includegraphics[width=0.58cm]{b10b.jpg} & 5.450 & 5.199 & 14.8\\ \hline
(xxiv) 11* \includegraphics[width=0.58cm]{b11a.jpg} &  6.297 & 6.016 & 93.2 \\ \hline
(xxv) 11 \includegraphics[width=0.58cm]{b11b.jpg} & 6.056  & 5.784 & 1.2 \\ \hline
(xxvi) 11 \includegraphics[width=0.58cm]{b11c.jpg} & 5.934  & 5.707 & 3.0 \\ \hline
(xxvii) 11 \includegraphics[width=0.58cm]{b11d.jpg} &  5.926 & 5.683 & 0.7 \\ \hline
(xxviii) 11 \includegraphics[width=0.58cm]{b11e.jpg} & 5.705  & 5.499 & 1.9 \\ \hline
(xxix) 12* \includegraphics[width=0.58cm]{b12a.jpg} & 6.962 & 6.684 & 29.7\\ \hline
(xxx) 12\ \, \includegraphics[width=0.58cm]{b12b.jpg} & 6.846 & 6.598 & 24.3\\ \hline
(xxxi) 12\ \, \includegraphics[width=0.58cm]{b12c.jpg} & 6.827 & 6.589 & 0.2\\ \hline
(xxxii) 12\ \, \includegraphics[width=0.58cm]{b12d.jpg} & 6.810 & 6.575 & 45.4 \\ \hline
(xxxiii) 12\ \, \includegraphics[width=0.58cm]{b12e.jpg} & 6.738 & 6.507 & 0.2\\ \hline
(xxxiv) 12\ \, \includegraphics[width=0.58cm]{b12f.jpg} & 6.514 & 6.324 & 0.2\\ \hline
\end{tabular}

\caption{Binding energies for $N$-solitons and binary species particles
(with frequency \%). \\ * denotes the global maximum of the binding
energy for each value of $N$.}
 \label{tab-binding}
\end{table}

To compute local maxima of the particle binding energy we use a simple
multi-start hill climbing algorithm with 1000 starts for each 
particle number $N,$ where each start consists of $N$ random positions
for the particles inside the square $[-20,20]\times[-20,20]$. 
The results for $2\le N\le 12$ are displayed in Figure \ref{fig-particles}
by plotting blue and red discs, centred at the positions of the blue
and red particles, with each disc diameter equal to the value of 
the separation $R$ at which the out of phase interaction potential,
$U_\pi(R)$, is minimized.
 
A total of 34 local maxima of $\delta$ were obtained from these 11000 
random starts and all are shown in Figure \ref{fig-particles}, ordered
first by increasing particle number $N$ and then by decreasing 
particle binding energy $\delta$ within solutions with the same value of
$N.$ For each $N$ the configuration with the global maximum of $\delta$ is
denoted by a *, and a comparison of these configurations with the soliton
solutions in Figure \ref{fig-solitons} reveals that the particle model
predicts the correct arrangement in each case. Furthermore, all of the
local maxima of the particle binding energy have an associated
static $N$-soliton solution that is a local minimum of the 
field theory energy. These soliton solutions have been obtained by 
using the particle results to specify the initial positions and phases
of the constituent single solitons used to generate initial conditions
for the field theory energy minimization algorithm.    

For all 34 configurations the value of the particle binding energy $\delta$
is listed in Table \ref{tab-binding}, together with the soliton binding
energy $\Delta$ of the associated $N$-soliton solution. 
The first column in this table denotes the configuration by reference to
the labelling in Figure \ref{fig-particles}. We see from this table that
the particle binding energy is a reasonable approximation to the soliton
binding energy, although it systematically underestimates the soliton value.
More importantly, there is an exact agreement between the particle model
binding energy ordering and that of the soliton model. Thus the particle
model correctly predicts the form of all the minimal energy solitons, 
even though it is a very simple model that includes only naive pair
interactions.  

A great advantage of the particle model is that it is computationally 
inexpensive to compute critical points of the binding energy, in 
comparison to more costly field theory simulations. In particular, this
allows a large number of simulations to be performed to provide a
measure of the capture basin for each critical point, and hence an
estimate of the likelihood of finding the associated soliton solution
in field theory computations, where the number of simulations that can
be performed with reasonable resources is quite low. 
The frequency percentage shown in the last column of Table \ref{tab-binding}
provides the percentage of the 1000 computations in the particle model
that produced the given critical point of $\delta.$
This data reveals that there are several examples where the optimal
configuration is not the one that is obtained most often. This already
occurs at the low value of $N=4,$ where more than three-quarters of the
simulations produced the linear chain, rather than the optimal square
arrangement. 

In typical multi-dimensional soliton investigations, 
it is not practical to perform more than a handful of simulations
for a given soliton number. In aloof soliton systems it therefore
appears quite likely that global minimal energy solitons may be
missed without the benefit of a reliable point particle model.
This situation is highlighted by the $N=9$ example, where the optimal
$3\times3$ square arrangement was not obtained from any of six field
theory simulations with random soliton initial conditions
(this explains the earlier comment that most, rather than all, 
 of the soliton solutions were obtained from randomly placed initial
solitons). 
Given the particle model results this is not surprising, 
as this optimal $N=9$ configuration was obtained from only 
$8.2\%$ of the 1000 random starting
configurations. There is therefore a danger that, in aloof theories,
soliton simulations alone may easily miss the optimal solution as capture
basins can be relatively small.

\section{A soliton lattice}\quad
The results presented so far with $N\le 12$ reveal that hexagonal and 
square constituents are the prevalent sub-structures. This suggests that
the optimal lattice (a doubly periodic configuration) will consist of  
either a hexagonal or a square arrangement. An indication that a
hexagonal arrangement is more favourable than a square one is provided by
the $N=16$ case.
\begin{figure}[ht]\centering
\hbox{
\ \hskip 0.3cm \includegraphics[width=6cm]{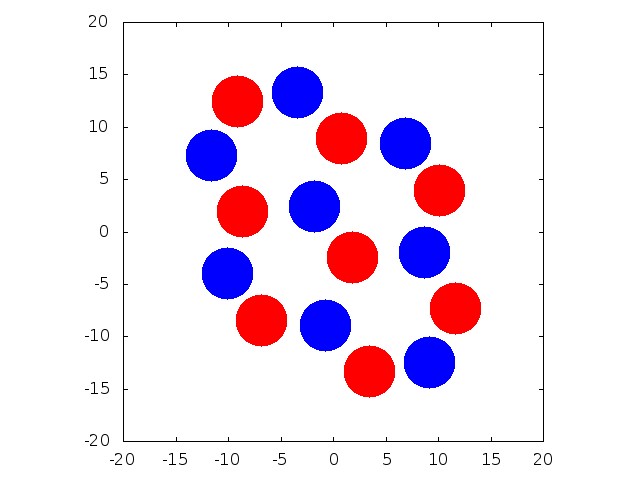}
\ \hskip 1.1cm \includegraphics[width=6cm]{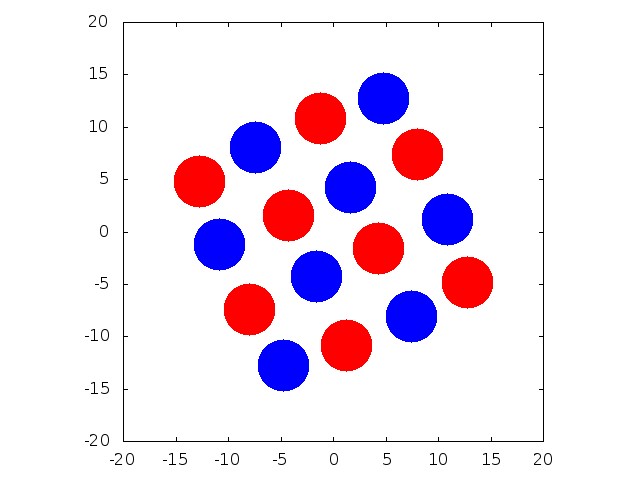}
}
\hbox{
\includegraphics[width=7.3cm]{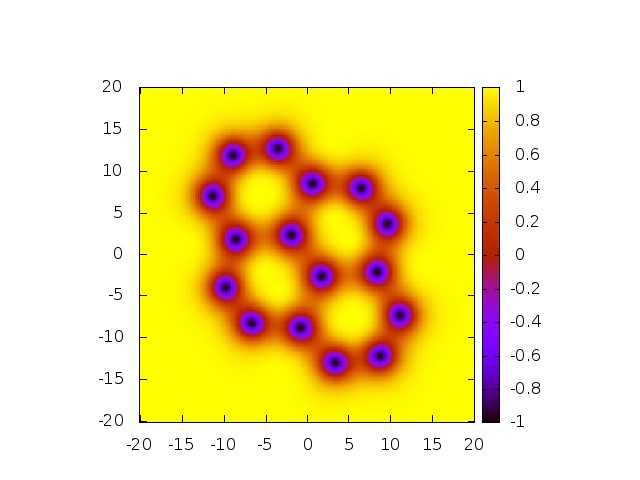}
\includegraphics[width=7.3cm]{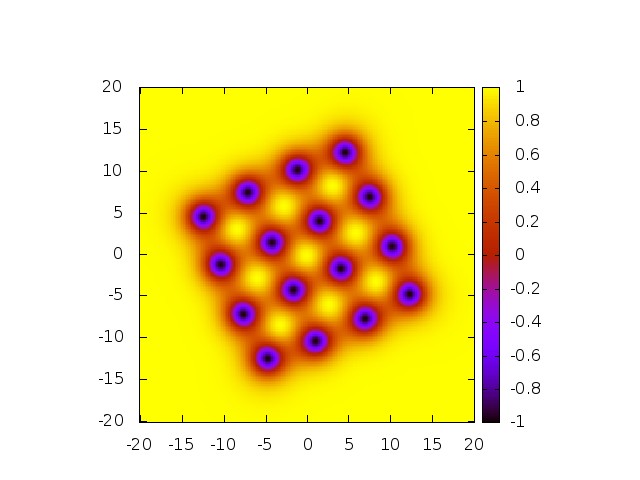}
}
\caption{$N=16$ critical points with hexagonal and square constituents. The top
row is the particle model and the bottom row 
displays $\phi_3$ for the field theory soliton solutions.}
\label{fig-B16}\end{figure}

The global maximum of the binding energy for 16 particles
is the hexagonal arrangement displayed in the top left image of
Figure \ref{fig-B16}. This configuration has binding energy 
$\delta=9.262$ and was obtained in 9.8\% of 1000 random starts.
The square configuration presented in the top right image of
Figure \ref{fig-B16} has a lower binding energy $\delta=9.184$
(it is not even the next best configuration after the hexagonal solution)
and was obtained in 6.8\% of the simulations. 
Using the particle
solutions to generate initial conditions for field theory simulations
generates the associated soliton solutions shown in the bottom row
of Figure \ref{fig-B16}. The soliton solutions confirm the energy
ordering of the particle model, with soliton binding energies of 
$\Delta=9.437$ and $\Delta=9.350$ 
for the hexagonal and square solutions respectively.

\begin{figure}[ht]\centering
\includegraphics[width=8cm]{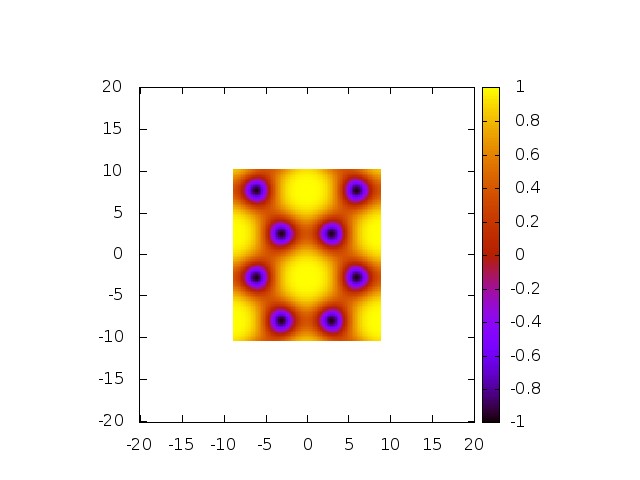}
\includegraphics[width=8cm]{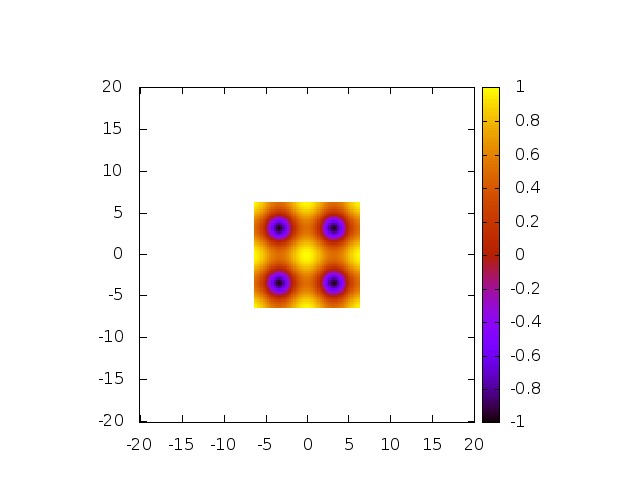}
\caption{Plots of $\phi_3$ in a periodic cell 
for the hexagonal lattice (left image) and the square lattice (right image).}
\label{fig-lat}\end{figure}

Guided by the hexagonal 16-soliton, we compute a doubly periodic 
hexagonal lattice containing four solitons on a fundamental
cell  
$(x_1,x_2)\in[-L/2,L/2]\times[-L/(2\sqrt{3}),L/(2\sqrt{3})]$ with
periodic boundary conditions. 
The maximal binding energy per soliton is $\Delta/N=0.678$,
 obtained when $L=18.03,$ at which the area per soliton is 
$L^2/(4\sqrt{3})=46.9.$ This soliton lattice is displayed in the
left image in Figure \ref{fig-lat}, where we show two fundamental
cells so that the hexagonal structure is clearly visible. 
The binding energy per soliton for this lattice is marked by the
blue line in Figure \ref{fig-bind} and is consistent with an 
asymptotic limit for the finite $N$ solutions computed earlier.
For large $N$ a candidate for the optimal $N$-soliton solution is
therefore an appropriate finite portion of this soliton lattice. 

The square lattice with a 
fundamental cell $(x_1,x_2)\in[-L/2,L/2]\times[-L/2,L/2]$ 
containing four solitons
has a maximal
binding energy per soliton of $\Delta/N=0.667$, obtained when $L=13.15$
with the area per soliton equal to $L^2/4=43.2.$ This confirms that the
hexagonal lattice is energetically preferred over the square lattice.
The right image in Figure \ref{fig-lat} displays
a fundamental cell for the square lattice.
The binding energy per soliton for this lattice is marked by the
red line in Figure \ref{fig-bind}.

We have verified that the conditions presented in \cite{JSS}, required 
for a doubly periodic solution to be a critical point with respect to 
variations of the lattice, are satisfied to a reasonable numerical accuracy.
Indeed, symmetry arguments and the results in \cite{Sp} show that both these
lattice solutions are not only critical points with respect to variations 
of the lattice but are also stable.

\section{Conclusion}\quad
Soliton binding energies in the standard Skyrme model are an order of 
magnitude larger than the nuclear binding energies that the model aims
to reproduce. Furthermore, the solitons are often too symmetric, due to
increased symmetries as individual solitons merge to form the bound state
solutions. 
The symmetry of the soliton is important as it relates directly to the 
allowed quantum states via a collective coordinate quantization \cite{FR}.
For example, the minimal energy 7-soliton has icosahedral 
symmetry \cite{BS2} with a lowest allowed spin state $J=\frac{7}{2}$
\cite{Kr2}, which disagrees with 
the ground state of $^7{\rm Li}$ with spin $J=\frac{3}{2}.$ 

A Skyrme model in which single solitons retain their individual identities
in bound states has the potential to resolve both the binding energy and 
symmetry issues. Here we have demonstrated that the required 
qualitative features can be obtained in a toy two-dimensional baby Skyrme
model, based on perturbing a repulsive limit of the model. This provides 
further motivation for current work in progress \cite{GHS} 
on the three-dimensional version of this problem.

We have demonstrated the utility of a binary species point particle
approximation. The binary species aspect arises because all solitons
can be assumed to have one of two possible internal phases. An alternative
point particle approach could involve a single species with an arbitrary
internal phase, but this generates the additional complication of 
computing the inter-particle force as a function of the relative 
internal phase. The results of the binary species model justify the use
of this approximation and renders the extra complexity unnecessary.
In a three-dimensional Skyrme model the internal phase is replaced by
an internal $SO(3)$ isospin orientation, so the point particle model is
a little more complicated. It remains to be seen whether a single species 
model, with each particle carrying an $SO(3)$ internal phase, 
or a multi-species model is more appropriate in this situation.

\section*{Acknowledgements}
We thank Mike Gillard, Derek Harland and Martin Speight for
useful discussions and a preview of their work in progress \cite{GHS}.
Our work is funded by the EPSRC grant EP/K003453/1 and 
the STFC grant ST/J000426/1.

\end{document}